\documentclass{emulateapj}
\slugcomment{{\sc Accepted to ApJ:} July 15, 2007}

\newcommand\um{\ensuremath{\mathrm{\ \mu{m}}}}

\newcommand\cm{\ensuremath{\mathrm{\ cm}}}
\newcommand\pc{\ensuremath{\mathrm{\ pc}}}

\newcommand{\LkHa}{LkH$\alpha$ }
\newcommand\Exp[1]{\ensuremath{\times10^{#1}}}
\newcommand\Msun{\ensuremath{\mathrm{M_\odot}}}
\newcommand\Rsun{\ensuremath{\mathrm{R_\odot}}}
\newcommand\Lsun{\ensuremath{\mathrm{L_\odot}}}

\newcommand\ang{\AA}

\newcommand{\SII}{[\ion{S}{2}] }
\newcommand{\FeII}{[\ion{Fe}{2}] }
\newcommand{\OI}{[\ion{O}{1}] }

\shorttitle{Bipolar Jet from \LkHa 233}
\shortauthors{Perrin and Graham}

\begin{document}

\title{Laser Guide Star Adaptive Optics Integral Field Spectroscopy\\ of a Tightly Collimated Bipolar Jet\\ 
from the Herbig Ae star \LkHa 233\altaffilmark{1}}

\author{Marshall D. Perrin and James R. Graham}
\affil{Astronomy Department, University of California,
    Berkeley, CA 94720-3411}
\email{mperrin@astro.berkeley.edu}

\altaffiltext{1}{Some of the data
presented herein were obtained at the W.M. Keck Observatory, which is
operated as a scientific partnership among the California Institute of
Technology, the University of California and the National Aeronautics
and Space Administration. The Observatory was made possible by the
generous financial support of the W.M. Keck Foundation.}

\begin{abstract}

We have used the integral field spectrograph OSIRIS and laser guide
star adaptive optics at Keck Observatory to obtain high angular
resolution ($0\farcs06$), moderate spectral resolution ($R\simeq
3800$) images of the bipolar jet from the Herbig Ae star \LkHa 233,
seen in near-IR \FeII emission at 1.600 \& 1.644 \micron. This jet is
narrow and tightly collimated, with an opening angle of only 9
degrees, and has an average radial velocity of $\sim 100$ km s$^{-1}$.
The jet and counterjet are asymmetric, with the red-shifted jet much
clumpier than its counterpart at the angular resolution of our
observations. The observed properties are in general similar to
jets seen around T Tauri stars, though it has a relatively large mass
flux of $1.2 \pm 0.3\Exp{-7} \Msun$ year$^{-1}$, near the high end of
the observed mass flux range around T Tauri stars.  We also spatially
resolve an inclined circumstellar disk around \LkHa 233, which
obscures the star from direct view. By comparison with numerical radiative
transfer disk models, we
estimate the disk midplane to be inclined $i= 65 \pm 5\degr$ relative to the plane of the sky. Since
the star is seen only in scattered light at near-infrared wavelengths,
we detect only a small fraction of its intrinsic flux. Because
previous estimates of its stellar properties did not account for this,
either LkHa 233 must be located closer than the previously believed,
or its true luminosity must be greater than previously supposed,
consistent with its being a $\sim4 \Msun$ star near the stellar
birthline.

\end{abstract}

\section{Introduction}

\begin{figure*}[t!]
\begin{center}
\includegraphics[width=6in]{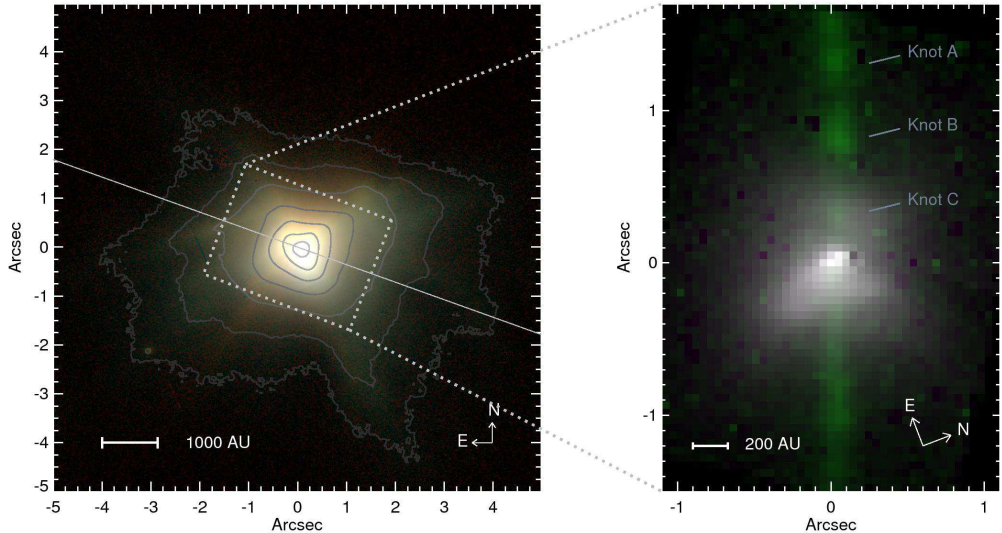}
\end{center}
\caption{
\label{diskjetfig}
\textbf{Left:} This $J,H,K_s$
composite from the OSIRIS imager shows the X-shaped
envelope surrounding \LkHa 233. The dashed rectangle shows the field
zoomed
in to on the right, while the solid line indicates the outflow axis.
\textbf{Right:} A portion of the
OSIRIS IFU data cube, centered in wavelength on the \FeII 1.644 $\micron$
 line (in
green) and rotated to show the jet axis vertical. This image reveals a tightly collimated bipolar outflow from
\LkHa 233 perpendicular to an edge-on circumstellar disk. The jet is
undetectable in the broadband image; only with the spectral
resolution of OSIRIS can it be detected. In the right hand panel, the
blue-shifted jet points down and the red-shifted jet is upward.
In these data we also see scattered light from a roughly edge-on
circumstellar disk and a larger bipolar nebula extending
perpendicular to the disk plane. The jet has excavated a cavity in
this envelope, the limb-brightened edges of which form the well-known
"X" centered on \LkHa 233  \citep{aspin85,Perrin2004Sci}.
}\end{figure*}

Bipolar outflows from young stars feature prominently in many 
of the most spectacular images of our universe
\citep[see,
e.g.][]{2007prpl.conf..215B}. But far from being merely aesthetically
pleasing, these jets play crucial roles in star formation. Molecular
cloud material must shed most of its angular momentum before it can
accrete onto a newborn star, and jets have been identified as a key
mechanism for this, removing angular momentum from disks and allowing
accretion to continue \citep[][ and references
therein]{2007prpl.conf..231R}.  Recent observations have for the first
time provided indications of jet rotation
\citep{2002ApJ...576..222B,2004ApJ...604..758C,2005A&A...432..149W,2007astro.ph..3271C},
suggesting that outflows may indeed carry angular momentum away
from their origin.  Jets have also been implicated in regulating the
overall efficiency of star formation by injecting turbulence into
molecular clouds \citep{2000ApJ...545..364M}.

The physical processes which drive these outflows remain poorly
understood. 
Several competing theories have been proposed to explain how gas is accelerated 
and collimated. The leading contenders all
invoke magnetic forces to shape the outflows, with the ``X-wind''
model \citep{Shang2007PPV} positing acceleration occurs near the star at the radius
of magnetospherical truncation, while the ``disk wind'' model
\citep{Pudritz2007PPV} posits acceleration from a wide range
of radii across the circumstellar disk. Alternatively, outflows could 
originate in 
spherical coronae, which are then collimated by surrounding magnetic
fields \citep{1999A&A...348..327S}. Outflows may even evolve through
each of these processes in turn \citep{2003Ap&SS.287...25S}.
Distinguishing between these theories will require detailed studies of
outflows as close to their origins as possible. At present, high
angular resolution observations of outflows have mostly focused on 
older T Tauri stars approaching the main sequence
\citep[see][and references therein]{2007prpl.conf..231R}. 
Observations of more sources at
younger ages are needed, especially considering that outflows are
intimately related to accretion, which is greatest at the
youngest ages \citep{1998ApJ...495..385H}.

In addition to age, we also need to understand how outflows vary with
stellar mass.  Because jets are thought to be driven by magnetic
fields, we might suppose that their properties should differ between
fully convective T Tauri stars and more massive stars which lack a
surface convection zone.  Indeed, outflows from high mass YSOs are
much less collimated than jets from T Tauri stars, with opening angles
of $30-60\degr$ \citep[e.g.
][]{1997ApJ...478..283H,1998ApJ...507..861S,2001Sci...292.1513S}.  The
Herbig Ae/Be stars \citep{her60,1998ARA&A..36..233W} are intermediate
between these two regimes, so studies of their outflows can
potentially clarify this transition in jet properties, which in turn
may shed light on the underlying physics through which jets are
produced. Yet relatively few outflows from Herbig Ae/Be stars have
been studied in detail, particularly at high resolution
\citep{1994ASPC...62..237M}.  In a very few cases, studies 
of parsec-scale outflows from Herbig Ae stars have
suggested that on large
spatial scales their velocities and collimation are largely
consistent with those observed in T Tauri jets
\citep{2004A&A...415..189M}.  

\begin{figure*}[t]
\begin{center}
\includegraphics[width=6.5in]{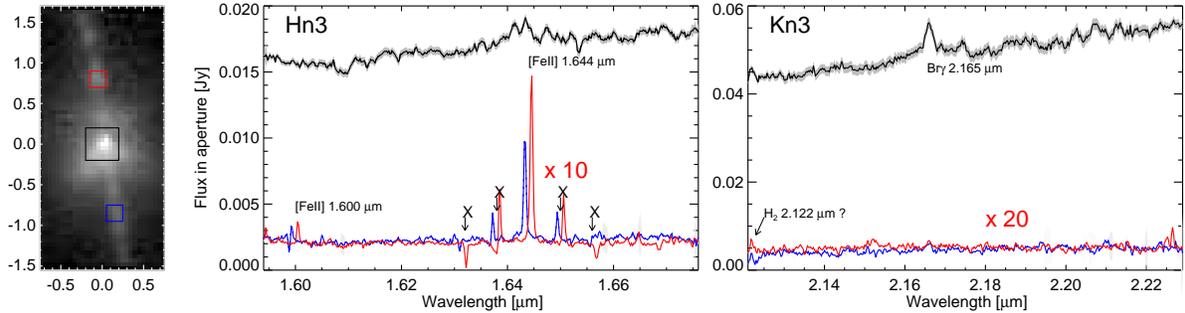}
\end{center}
\caption
{
\label{fig1d}
One dimensional spectra extracted from the OSIRIS data cube.
\textbf{Left:} Square apertures indicate the regions extracted to produce the spectra at right, 
shown in corresponding colors. The
red aperture is centered on knot B.
\textbf{Center and Right:} One dimensional spectra extracted from the
data cube for those apertures, in OSIRIS' $Hn3$ and $Kn3$ filters; the
light grey shading shows the statistical uncertainty. For clarity, the
emission line spectra have been scaled up $\times 10$ and $\times20$ relative to
the continuum. The outflow is clearly detected in both the
1.600 and 1.644 \micron\ lines of \FeII, and excess emission in
Br$\gamma$ is seen at the
location of the star. There is a hint of 2.122 \micron\ H$_2$ emission, but it unfortunately
falls at the very edge of our bandpass so its reality is uncertain. The
four features marked ``X'' are spectral ghosts caused by a grating misalignment
during OSIRIS commissioning, which has since been fixed. The
observed stellar continuum is $\sim 1/4$ the 2MASS flux levels
for \LkHa 233, reflecting the fact that our aperture does not include the whole
extended source, possibly stellar variability and/or uncertainties
in flux calibration. 
}
\end{figure*}

But a lack of high angular resolution studies of Herbig Ae star outflows 
has thus far prevented detailed 
comparison with outflows from T Tauri stars. For T Tauri stars, we now
have a sophisticated understanding of outflow properties on scales of a
few tens of AU, derived from high angular resolution observations. 
Because stellar outflows are detectable 
primarily in emission lines from shock-excited ions, line ratios
can provide measurements of many physical parameters within the
outflow. When both spatially and spectrally resolved data are
available, very detailed portraits of outflows have been obtained 
\citep[e.g.][]{1999A&A...350..917B,2004AJ....127..408B,2007ApJ...660..426H}.
Since outflows have characteristic spatial scales of a few
AU (corresponding to $0\farcs01-0\farcs1$ at typical distances) and
velocities of tens to hundreds of km s$^{-1}$, the necessary
observations combine high angular resolution ($< 0\farcs1$)
with moderate spectral resolution ($R \sim 2000-4000$).  Having both
spatially and spectrally resolved data is key to disentangling the
outflow structures as close to the disk as possible.  In optical
wavelengths, such as for the \SII 6717, 6731 \ang\ and \OI 6300, 6364
\ang\ emission lines, the necessary angular resolution is available
only from space
\citep[e.g.,][]{1999ApJ...512..901H,2000ApJ...537L..49B,2002ApJ...580..336W,2004ApJ...604..758C,2007ApJ...660..426H}
However, the arrival of integral field spectroscopy with adaptive
optics on 8-10 m telescopes has now opened the door to high angular resolution
studies of jets from the ground.  At the near-infrared wavelengths
suitable for adaptive optics, several emission lines of \FeII between
$1.2 - 1.6 \mu$m are ideal tracers of shocked outflows
\citep{1987ApJ...313..847G}.  Furthermore, these lines provide
diagnostic potential enabling the determination of physical conditions
within the jet \citep{2003A&A...410..155P,2004ApJ...614L..69H}, and
ultimately the mass flux, a key parameter in any model of outflow
physics.

In this paper, we seek to investigate whether the similarity on large
spatial scales between
outflows from T Tauri and Herbig Ae stars still
holds true on finer spatial scales. Toward that
end, we present the first high angular resolution integral field
spectroscopy of an outflow from a Herbig Ae star, enabling us to
investigate its collimation, kinematics, and certain physical
parameters on angular scales of $\sim 60$ milliarcseconds. By
comparison with the properties of jets seen from T Tauri stars, we
hope to clarify the influence of stellar mass upon outflow physics.
Also, from a technical perspective, we wish to demonstrate the utilty of
laser guide star adaptive optics (LGS AO) for
obtaining high angular resolution spectroscopy of an outflow from a
fairly faint and obscured source. Because many young stars which drive
jets are still deeply embedded in their dusty birthplaces, only a limited number
are accessible to traditional natural guide star AO. The
growing availability of LGS AO on large
telescopes promises to enable high angular resolution studies of
a greatly increased target sample, particularly at the youngest ages.

We present here observations made with Keck's new integral field spectrograph
OSIRIS, using laser guide star adaptive optics, which show in
exquisite detail a tightly collimated jet from the Herbig Ae star 
\LkHa 233 (also V375 Lac, HBC 313). This star, part of the original Ae
star sample identified by \citep{her60}, is 
one of the relatively few Herbig Ae/Be stars known to possess a jet.
It is an A4 star, with
$\log T_{eff} = 3.93$ \citep{2004AJ....127.1682H}.
Its distance has generally been considered to be 
 880 pc \citep{1978MNRAS.182..687C}.
We adopt this distance as well, but some authors have
argued for a closer distance;  see \S \ref{distance} below.
Hernandez et al.\ also derived $M_* = 2.9~
M_\odot$, $L_* = 80~ L_\odot$, and age = 2.61 Myr, with $A_V$ = 3.7, $R_V =
5$, but those values are likely not accurate, because 
they were derived without accounting for the fact that
\LkHa 233 is seen only in scattered light. Hence both 
the extinction to the star and its total luminosity must be higher; see \S
\ref{disksection}.

\LkHa 233 sits at the center of an
X-shaped bipolar nebula visible from optical to near-infrared
wavelengths \citep{aspin85,Li94}. Near infrared polarimetry shows a
dark lane passing in front of the star, indicating the presence of a roughly
edge-on circumstellar disk \citep{Perrin2004Sci}.
\citet{corc98} discovered that \LkHa 233 possesses an extended bipolar
jet and counterjet visible in the optical \SII 6716, 6731 \ang~lines. The
blueshifted jet extends toward position angle (PA) 250\degr, perpendicular to the inferred circumstellar disk 
and along the symmetry axis of the
surrounding nebula.
The redshifted counterjet, at 70\degr, was not detected within 0\farcs6 of
the star, which Corcoran et al.~interpreted as occultation by
the circumstellar disk. A followup study by
\citet{2004A&A...415..189M}
found that a chain of Herbig Haro objects
extends almost 2 pc to either side of \LkHa 233, located
approximately along the jet axis but with considerable scatter.

In the following section, \S 2, we present our observations and
outline the data reduction process. We then describe in \S 3 the observed
properties of both the jet and the circumstellar disk, derive
estimates for the mass flux in the jet, and re-examine the question of
the distance to \LkHa 233. In \S 4 we discuss the implications of
these results, with special attention to comparison with the observed
properties of jet around other Herbig Ae and T Tauri stars.
Conclusions and prospects for future work are given in \S 5.

\section{Observations}

We observed \LkHa 233 on 2005 Oct 13 UTC using the integral field
spectrograph OSIRIS and laser guide star
adaptive optics system \citep{2006PASP..118..297W,2006PASP..118..310V} on the W. M. Keck II telescope. 
The performance of the 
\LkHa 233 itself served as the tip-tilt
reference for the adaptive optics system. The skies were clear and
seeing was average, $0\farcs6$ at $K'$, with winds from 15-20 km hr$^{-1}$.

OSIRIS is a lenslet-based integral field spectrograph which provides a spectral resolution of 
3800 from 1-2.5 microns \citep{2003SPIE.4841.1600L}. We selected its
50 mas pixel$^{-1}$ spectrographic plate scale for our
observations. While this scale does not Nyquist-sample the
diffraction-limited PSF, it provides a
larger field of view and better signal to noise per integration time
for extended emission
than the finer plate scales. We observed \LkHa 233 using the narrow-band \textit{Hn3}
and \textit{Kn3} filters, covering the wavelength ranges
1.594-1.676 and 2.121-2.229 $\micron$, respectively. Total integration
time on-source was 1800 s for \textit{Hn3} and 900 s for \textit{Kn3}. Between exposures
on \LkHa 233, we obtained equally deep sky observations at a position 20"
away.

\begin{figure*}[t]
\begin{center}
\includegraphics[width=6in]{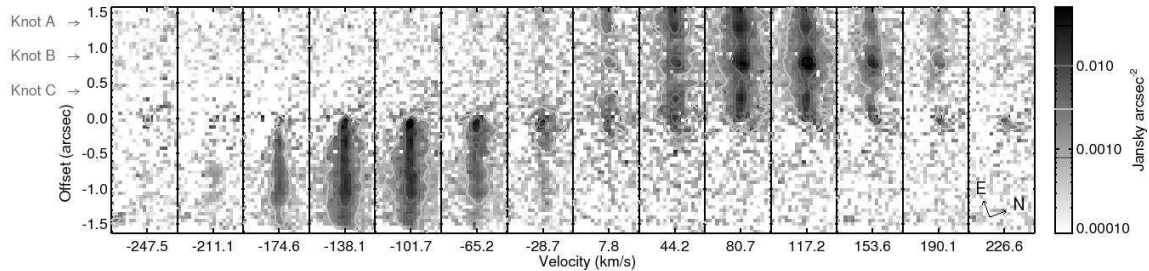}
\end{center}
\caption{
\label{slicesfig}
Continuum-subtracted velocity channel maps for \FeII 
1.644 \micron\ emission from the jet. Each panel shows the
intensity image for a given velocity (relative to the solar 
system barycenter) noted at the bottom of each panel.
These are contiguous spectral channels from OSIRIS, 
and the field of view shown is $0\farcs9 \times 3\farcs2$. 
The blueshifted (bottom) jet is smooth and relatively featureless in both 
intensity and velocity, while the redshifted (top) jet breaks into
three distinct clumps, with the central clump having a higher velocity
than the other two.
}
\end{figure*}

We reduced the data using an early version of the OSIRIS data
reduction pipeline \citep{2004SPIE.5492.1403K}.  The data were
sky-subtracted, corrected for offsets between the 32 detector readout
channels, spatially rectified to extract the individual spectra into a data cube, 
and corrected for cosmic rays and other
blemishes.  The A stars BD +9 5007 and HD 7789 served as telluric
standards and Point Response Function\footnote{We follow the Spitzer convention of 
distinguishing between the intrinsic optical Point Spread Function (PSF) and
the Point Response Function (PRF), which is the PSF convolved with the
detector's pixel response. See 
\url{http://ssc.spitzer.caltech.edu/postbcd/doc/PRF\_vs\_PSF.pdf}.
Because our OSIRIS data used the 50 mas pixel scale, coarser than
Nyquist for the Keck AO PSF at 1.6 \micron, the
PRF is dominated by the pixel response.  By taking the 
quadrature difference of the observed 61 mas PRF FWHM and 50 mas pixel scale, 
we estimate that the intrinsic optical PSF had $\sim 35$ mas FWHM, 
approximately the diffraction limit. Because of the difficulties inherent in trying to 
accurately measure Strehl ratios on undersampled
data \citep{2004SPIE.5490..504R}, we do not attempt to
directly measure the Strehl ratio achieved in these observations.}
(PRF) references. For BD +9 5007, observed immediately
before \LkHa 233, we measure a PRF FWHM of $61 \pm 5$ mas.
We also used these stars to establish photometric zero points: 
we converted their 2MASS $H$ and $K_s$ magnitudes into magnitudes in the OSIRIS \textit{Hn3} and \textit{Kn3} filters via
synthetic photometry, and then derived from those magnitudes a zero point for flux-calibrating our
\LkHa 233 data.

These observations were made during the
instrument's commissioning period, and show a variety of instrumental
artifacts, most notably ghost images at certain wavelengths caused by
crosstalk between lenslets due to a grating misalignment.
These ghost images do not occur in the wavelength ranges of interest
to us, and so we simply ignore them.  The grating has
been realigned subsequent to these observations and more recent
data do not suffer from these ghosts.

OSIRIS also includes an imaging
camera whose field of view is offset by 19\farcs4 from the spectrograph
field. Our sky observations were obtained such that \LkHa 233 fell on
the imager field of view (FOV) while we were obtaining spectrograph skies, and vice
versa.  In this manner we obtained $Z, J, H,$ and $K$ band observations of \LkHa 233, with
total exposure times of 300, 360, 360, and 315 s, respectively. These
observations were reduced via the usual steps of flat fielding, sky
subtraction, image registration, and summation.

During this observing run, we also observed the Herbig Ae/Be stars
\LkHa 198, V376 Cas, and Parsamian 21, with identical instrument
configuration and exposure times. No outflows were visible around the
first two sources. Parsamian 21 has a faint outflow extending
to the north, which we postpone discussing for a subsequent paper.

\section{Results}

\subsection{The Jet}
\begin{figure*}[!ht]
\begin{center}
\includegraphics[width=6in]{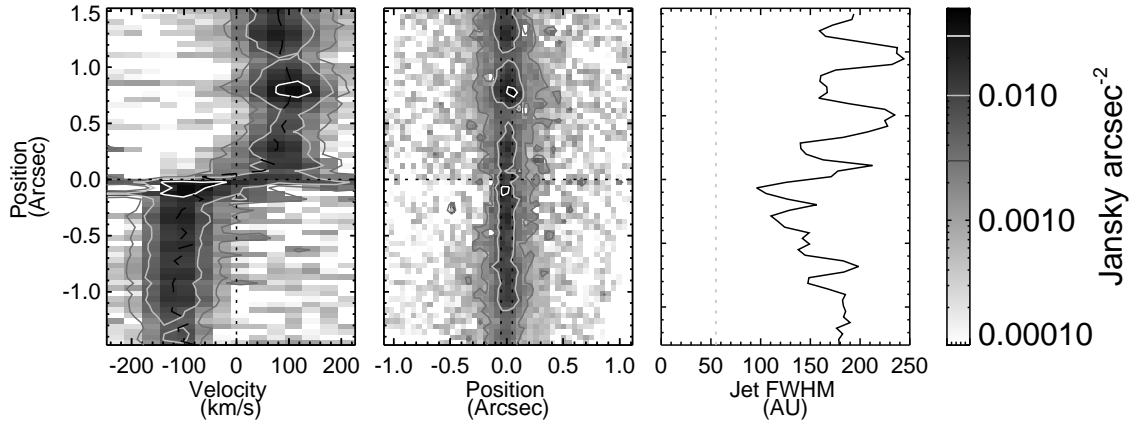}
\end{center}
\caption{
\label{plotfig}
\textbf{Left:} Continuum-subtracted position velocity diagram for
\FeII 1.644 \micron, using a
0\farcs1 wide ``virtual slit'' centered on the jet. 
\textbf{Center:} Continuum-subtracted image of the jet, integrated in
velocity. The upper and
lower portions of this image are summed over slightly different
wavelength ranges centered on the red- and blueshifted jets, each with
bandwidth 0.6 nm. The vertical dotted lines show the boundary of the ``virtual
slit'' used to produce the left-hand figure.
\textbf{Right:} Measurements of jet FWHM versus distance from the
star. This plot was constructed by fitting each row in the
continuum-subtracted image with a Gaussian. The FWHM appears narrower
at the position of the knots, and broader where the emission is
fainter (see \S \ref{jetsection}).
}
\end{figure*}

\label{jetsection}

A well-collimated bipolar emission-line jet extends several arcseconds
on either side of \LkHa 233 (Figure \ref{diskjetfig}), with the
blueshifted jet at a position angle of 249\degr, and the redshifted
jet at 69\degr.  We also see a roughly edge-on circumstellar disk and
bipolar nebula extending perpendicular to the jet. The jet has
excavated a cavity in this envelope, the limb-brightened edges of
which form the well-known ``X'' centered on \LkHa 233.  We detect the
jet in the \FeII 1.600 and 1.644 $\micron$ lines, and possibly H$_2$
2.122 \micron\ (Figure \ref{fig1d}).  Faint hydrogen Br$\gamma$ 2.16
$\micron$ emission is also detected along the course of the jet, but
does not show velocity shifts and is spread over a wider spatial
extent than the [FeII] emission; this appears to just be Br$\gamma$
emission from the star, scattered by dust in the circumstellar
envelope.  Of the lines observed here, the 1.644 $\micron$ line is the
brightest, and we choose to concentrate our analysis on it. In Figure
\ref{slicesfig} we show continuum-subtracted images in a range of
velocity channels centered on 1.644 $\micron$, and we show in Figure
\ref{plotfig} a position-velocity diagram, an summed
continuum-subtracted jet image, and a plot of the FWHM.  

The jet is narrow, tightly collimated, and knotty. Its FWHM increases
slowly from 100 AU near its base to $\sim 200$ AU at the edge of our
FOV, 1300 AU from the star, giving an opening angle of 9\degr (Figure
\ref{plotfig}).  Three bright knots are apparent in the redshifted
jet, at approximate distances of 0.3, 0.8, and 1.3\arcsec\ from the
star, corresponding to 260, 700, and 1150 AU in the plane of the sky.
We will refer to these as knots C, B, and A, respectively.  The
blueshifted jet is noticeably less clumpy, but brightens considerably
near its base, reaching its peak intensity at 0\farcs1 (88 AU) from
the star.
There is an anticorrelation between jet width and intensity (Figure
\ref{plotfig}, right), with the bright knots having lower FWHM than
the fainter inter-knot emission. This behavior has also been observed
for the outflows from RW Aur (Woitas et al. 2002) and HH 23 (Ray et
al. 1996). 

We were unable to find any reported measurements of \LkHa 233's radial
velocity in the literature\footnote{We note that Corcoran and Ray
(1997,1998) state they measured the systemic velocity using optical Na
D lines, but do not anywhere report the actual measured velocity}, nor
are there stellar lines suitable for measuring it in our data (since
$Br\gamma$ is in emission). Hence we compute velocities relative to
the solar system barycenter, without correcting for the radial motion
of the star.  In this reference frame, the red- and blueshifted jets
have average velocities of $+108 \pm3$ and $-89 \pm 3$ km s$^{-1}$,
respectively.  If the red and blue jets have the same speed relative
to the star but opposite directions, then the stellar barycentric
radial velocity appears to be $\sim -10$ km s$^{-1}$.

A significant difference between the two directions is that the
blueshifted jet has a relatively constant velocity along its length,
while the redshifted jet shows variation in its velocity.  Knot B, at
$0\farcs8$, has a higher velocity than the other two clumps by about
30 km s$^{-1}$ (Figure \ref{plotfig}, left). This velocity variation
so close to the jet's origin indicates that the amount of acceleration
imparted to the jet must fluctuate with time, and must do so in a way
that is not symmetric between the jet and counterjet.

If we assume the jet is perpendicular to the disk, which has an estimated inclination of $i = 65\degr \pm 5$ (see \S\ref{disksection} below),
then the radial velocities we
observe are only $40 \pm 9$ percent of the total jet velocity. The implied total jet
velocity, 200-300 km s$^{-1}$, is consistent with outflows seen
around other Herbig Ae stars \citep{1994ASPC...62..237M}.  Based on
this, we estimate the
proper motion of the jet should be $0\farcs05\pm0.01$ per
year. Hence the bright knots in the jet were launched from the star
within the last 10 to 20 years, and their motion should be apparent
over short timescales.

We do not see much evidence for the low velocity component (LVC)
reported by \citet{corc98} at about -25 km/s. This may be due to the fact that
the \FeII 1.644 $\micron$ line traces higher densities than the \SII
6716, 6731~\AA~ lines, so in the NIR we are only sensitive to dense
gas in the core of the jet.

\begin{figure}[!t]
\begin{center}
 \includegraphics[width=2in]{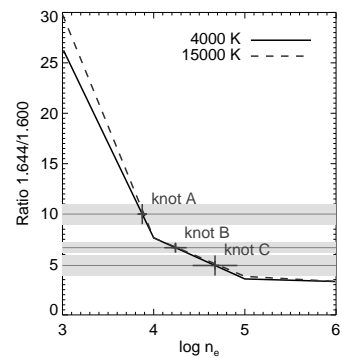}
\end{center}
\caption{
\label{electronfig}
The ratio of the 
 1.644 \um\ and 1.600 \um\ lines provides a
 diagnostic for electron density $n_e$. The solid and dashed lines
 show the expected ratio as a function of $n_e$ for two different
 temperatures, from Nisini et al. (2002). The measured line ratios for the
 three clumps in the redshifted jet of \LkHa 233 are indicated; the
 length of the crosses shows the uncertainties. The electron density
 decreases along the jet as one moves away from the star. The values
 of $\sim 10^4- 10^{4.5}$ cm$^{-3}$ are comparable to electron densities 
 in several T Tauri HH jets measured from the same \FeII line ratio by
 \citet{2002A&A...393.1035N,2005A&A...441..159N}.
}
\end{figure}

The emission lines of \FeII can be used to measure several important
physical quantities, such as temperature, electron density,
and extinction
\citep{2004ApJ...614L..69H}. Most such
measurements require the comparison of flux ratios between different
lines of \FeII, or between \FeII and other species in the optical. 
The limited
wavelength coverage of our data restricts their
diagnostic capabilities, as only the 1.644 and 1.600 $\mu$m lines
are present in our wavelength range. The flux ratio of those
two lines does provide a measure of the electron density in the shocked gas
\citep{2002A&A...393.1035N}. We find that the electron density in \LkHa 233's
jet is $n_e \sim 10^4-10^{4.5}$ cm$^{-3}$ and decreases at
larger distances from the star; see Figure \ref{electronfig}. This
decrease at larger distances has also been seen in many jets from T
Tauri stars 
\citep{2002A&A...393.1035N, 2006A&A...456..189P, 2007AJ....133.1221B, 2007ApJ...660..426H}
and is believed to show the slow recombination of hydrogen after the
initial shock ionization \citep{1999A&A...342..717B}.
However, as \FeII emission traces
denser regions of the gas, our measured electron densities 
are not
necessarily indicative of \LkHa 233's outflow as a whole. Future observations
which also include the optical \SII doublet are needed to provide complementary
measures for gas at lower densities and/or different distances from
the shock fronts.

\subsection{The Disk}
\label{disksection}
\begin{figure*}[!ht]
\begin{center}
\includegraphics[width=6in]{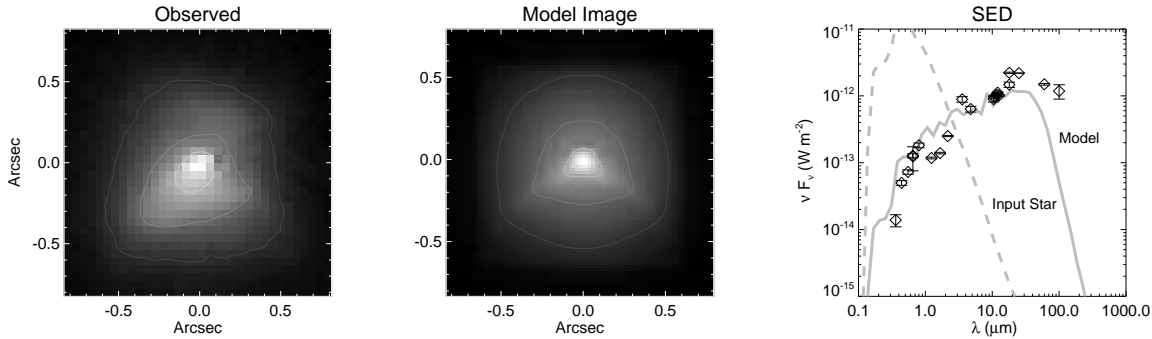}
\end{center}
\caption{
\label{modelfig}
\textbf{Left:} Observed appearance of the disk around \LkHa 233. This
$1.6 \um$ image is the median of the continuum from the OSIRIS
\textit{Hn3}
datacube.
\textbf{Center:} A $1.6 \um$ disk model computed with MCFOST,
convolved with a measured OSIRIS \textit{Hn3} PRF obtained from observations
of a calibration star. The model reproduces the central brightening,
sharp V-shaped southern disk face, and less distinct northern disk face.
\textbf{Right:} The observed SED for \LkHa 233 (diamonds), plus the
computed SED from our model (solid line).  The SED is fairly well fit for
wavelengths $< 50 \um$; the deficit at longer wavelengths arises
because the model presented here includes only the disk and not the
surrounding extended nebula, which contains cooler
dust at much larger distances from the star (but still within the
large IRAS 60 and $100 \um$ beams). The dashed line is the input
stellar spectrum, and the solid line shows the resulting spectrum
after reprocessing by the disk. See \S 3.2 in the text for a
discussion of this model and its limitations.
}
\end{figure*}

In addition to showing the jet, these observations also clearly reveal
the circumstellar disk around \LkHa 233, in the form of a
bright nebula crossed by a dark lane, the
characteristic signature of a roughly edge-on circumstellar disk
(Figure
\ref{diskjetfig}, right).  The presence of this disk was originally
suspected based on optical polarimetry \citep{aspin85}.
\citet{corc98} were unable to trace the redshifted optical \SII
emission any closer than 0\farcs6 to the star, and inferred the
presence of a disk with this radius that occulted the jet. We confirm here that
interpretation: the circumstellar disk extends from $0\farcs5$ below to
$0\farcs6$ above the locus of peak intensity. In contrast to the \SII
line, in \FeII 1.644 $\micron$ light the jet can be traced all the way in
to the star on both sides (Figure \ref{slicesfig}). There is at most a
two-pixel-wide ($0\farcs1$) region at the base of the red jet where
the emission is somewhat fainter, possibly due to obscuration. This
indicates that the circumstellar dust is not optically
thick at these wavelengths, in contrast to the optical. The observed
disk morphology is consistent with that seen in visible light in HST
WFPC2 observations (Stapelfeldt et al., in prep).

The continuum intensity peak is well-resolved in these data, with
a FWHM of 0\farcs135 versus 0\farcs06 for a PRF reference star.
Therefore we do not observe the star directly, but instead see it in
disk-scattered light.  The peak intensity is located at the apex of
the southwest face of the disk, which is the brighter of the two
faces. This is consistent with the southwest side of the system being
inclined toward us, as indicated by the blueshifted jet.
By comparison with a series of circumstellar disk models computed
using the MCFOST Monte Carlo radiative transfer code of
\citet{2006A&A...459..797P}, we determined that \LkHa 233's disk is inclined
approximately 
$65 \pm 5$ degrees\footnote{We measure inclinations of the disk midplane relative to the plane of the sky, such that pole-on disks have $i=0\degr$ and edge-on disks have $i=90\degr$}. A numerical model of a 
circumstellar disk
at this inclination extending from 0.6 to 500 AU reproduces fairly well both the 
observed appearance and SED of \LkHa 233 (Figure \ref{modelfig}). We
caution that some parameters, such as the inner disk
radius, are not well constrained by these data. Hence
this is only one possible disk model which fits the
current data,
not necessarily a unique best-fit solution.

Previous estimates of the stellar properties of
\LkHa 233 itself were obtained assuming only regular
interstellar extinction, typically around $A_V=2-4$, with reddening $R_V$ between
3 and 5
\citep{2004AJ....127.1682H,2006astro.ph..8541M}.
However, our results here confirm that \LkHa 233 is seen entirely in
scattered light at optical and near-IR wavelengths. The true $A_V$ is much higher than previously estimated, 
so the observed SED contains only a small
fraction of the total luminosity. \LkHa 233 must therefore be
substantially more luminous than previously thought. With the
stellar temperature fixed at $T_\ast = 8600 K$ (appropriate for an A5
star), we were able to
simultaneously match both the observed disk morphology and SED by
using a stellar radius of $R_\ast = 9 \Rsun$. This implies a total
stellar luminosity $L_\ast \sim 400 \Lsun$---about an order of
magnitude higher than the 20-80 \Lsun previously found by
\citet{2004AJ....127.1682H}.
These revised stellar properties place \LkHa 233
near the stellar birthline for intermediate mass stars
\citep{1990ApJ...360L..47P,2005IAUS..227..196P}, and therefore \LkHa
233 may in fact be an very young star with mass $\sim 4 \Msun$.

The key parameters of our model are as follows: $T_{\ast}$ = 8600 K,
$R_\ast$ = 9 \Rsun, $r_{out} = 500$ AU, and
grain sizes from $0.03-10 \um$ with power law
index -3.5. 
However, we emphasize that the model presented here is \textit{not} a rigorous
best fit to the data, merely one plausible fit, and there remain large uncertainties in
both disk and stellar properties.  In particular, the MCFOST Monte Carlo code
used does not yet account for any puffed up inner rim to the disk
\citep{2004A&A...417..159D} nor does it include the extended bipolar
envelope also present around \LkHa 233 in addition to the disk \citep{Perrin2004Sci}.
The lack of the bipolar envelope component results in the present
model having an unrealistically large scale height (30 AU at 100 AU)
in order to fit the data well. 
More detailed modeling, quantitatively fit to observations
across a wide range of wavelengths, would help clarify the situation,
but is beyond the scope of this paper.

\subsection{Mass Flux}

The observed intensity in the \FeII lines allows us to estimate the mass flux.
Currently the physical parameters of the jet, such as the temperature,
ionization, and Fe depletion onto grains, are not well known so our
estimate will require making several assumptions; future
observations will enable a more refined calculation.
\citet{2006A&A...456..189P} presented two separate methods for
estimating mass flux based on \FeII emission, based on either 
(A) the jet's apparent size and density, or (B) the observed emission line
luminosity.

The first method relies on computing the mass outflow rate 
across a given plane perpendicular to the jet, 
based on
the observed jet cross section, velocity, and density: 
\[
\dot M = \mu m_H n_H \pi r_J^2 v_J
\]
where $\mu$ is the mean molecular weight, taken to be 1.24, $m_H$ is
the mass of hydrogen, $n_H$ is the number density of hydrogen, and
$v_J$ is the jet velocity.
In the case of \LkHa 233, we do not know $n_H$, but
we do know that $n_e \sim 5\Exp{4} \cm^{-3}$, from which we can estimate $n_H$. 
Typical
ionization fractions for HH objects are $x_e = 0.03-0.3$, as determined from both
shock models and observational diagnostics
\citep{1994ApJ...436..125H,1999A&A...342..717B,2000A&A...356L..41L}, so if we assume
$x_e = 0.1$, then $n_H = 5\Exp{5} \cm^{-3}$. Using a rough estimate of
$r_J = 90$ AU, we compute $\dot M = 1.0\Exp{-6} \Msun$ year$^{-1}$.
This method relies on the assumption that the jet can be treated as a
uniformly filled cylinder with density measured by our diagnostic, and thus
may be better considered an upper limit on the true $\dot M$.

The second method relies on the observed luminosity $L_{[FeII]}$ being
proportional to the total number of emitting atoms.
Unlike the first method, this second approach does take into account spatial
variations in the jet density, but is subject to uncertainties in the
absolute flux calibration. In this method we compute $\dot{M}$ using
\begin{eqnarray*}
    \dot M &=& \mu m_H (n_H V) v_t/l_t \\
    (n_H V) &=& L_{[FeII]} \left( h \nu A_i f_i \frac{Fe+}{Fe}
    \frac{Fe}{H} \right)^{-1}
\end{eqnarray*}
where $V$ is
the volume filled by the emitting gas, $v_t$ is the tangential
velocity of the jet, $l_t$ is the tangential length of the jet volume element, 
$\case{Fe+}{Fe}$ and $\case{Fe}{H}$ are the ionization fraction and
atomic abundance relative to hydrogen, 
$A_i$ is the Einstein A radiative rate coefficient, and $h$ and $\nu$
are as usual the Planck constant and frequency of the radiation.
Based on our estimated disk inclination $i=65\degr\pm5$ and observed
$v_{radial} = 100$ km s$^{-1}$, we compute $v_t \sim 240\pm40$ km s$^{-1}$, and
we take $l_t$ to be the tangential length of each OSIRIS pixel,
$44$ AU at 880 pc.  
Several assumptions are needed 
in order to transform the
observed line luminosity $L_{[FeII]}$ into the density-volume product
$n_H V$: without accurate knowledge of the temperature we have to
estimate the ionization fraction and upper state fraction for iron. We 
base our estimates on conditions observed in typical HH 
objects, for instance by \citet{2006A&A...456..189P} and \citet{2007ApJ...660..426H}, who measured temperatures 
of $8000-25000$ K based on optical line ratios.
Taking those temperatures plus our measured $n_e=10^4 - 10^{4.5}$, we
used a 16-level model of the Fe+ ion \cite[][similar to that described by]{2003A&A...410..155P}
to estimate the fraction of ions in the upper energy level is $f_i = 0.01 \pm 0.005$.
We further assumed
that Fe is entirely ionized ($\frac{Fe+}{Fe} = 1$), and that iron
has a solar abundance but is 90\% depleted onto solid grains (based on
the 87\% depletion reported for the HH 23 jet by Podio et al. 2006).

Our spatially resolved IFU data enable us to measure the mass flux rate 
as it varies along the jet. 
We first collapse the  2d continuum-subtracted \FeII image perpendicular to the jet to get a total $L_{[FeII]}$ for each row, and 
then apply the above algorithm to compute $\dot M$.
The computed mass flux has a median value of
$1.2\Exp{-7} \Msun$ year$^{-1}$, and is fairly constant along most of
the jet, varying between $1.0-1.5\Exp{-7} \Msun$ year$^{-1}$. The one
exception is the middle knot in the redshifted jet, knot B, where the flux
rises to $2.6\Exp{-7} \Msun$ year$^{-1}$. 
We caution that these values are \textit{approximate} at best, given
the lack of knowledge of the temperature and ionization structure, and
furthermore represent lower limits, because all of the outflow is not
necessarily shocked and luminous, and because we have not corrected for
extinction.

The two methods give results which differ by an order of magnitude, with the 
estimate based on method A higher than that of method B.
This discrepancy is not entirely unexpected,
since the filling factor for \FeII emission is less than unity.
Podio et
al.\ (2006) found masses computed with method A were often 3-10 times higher
than those computed based on emission line fluxes, and they considered
the latter method to provide a better estimate of the mass flux
transported by the atomic component of the outflow. 

The values for $\dot M$ for \LkHa 233 are comparable to the
high end of the range of mass fluxes seen around T Tauri stars. For
instance, if we adopt the lower of the two estimates as the more reliable, 
the flux in \LkHa 233's jet is
roughly comparable to the mass flux computed for RW Aur by Woitas et
al.\ (2002), $\sim 10^{-7} \Msun$ year$^{-1}$, though
about an order of magnitude higher 
than the values measured for HH 34 and HH 111 by Podio et
al.\ (2006).

Outflow rates from young stars are proportional to
the accretion rates onto those stars
\citep{1990ApJ...354..687C,1995ApJ...452..736H,2006ApJ...646..319E},
across a mass range spanning
from brown dwarfs \citep{2005ApJ...626..498M} to Herbig Ae/Be stars
\citep{1998A&A...331..147C}.
Outflow mass fluxes are typically a few percent of the accreting mass flux \citep{1995ApJ...452..736H}
As far as we know, there are currently no measurements of the accretion rate onto \LkHa
233, but based on our estimated outflow mass flux, the accretion rate of onto it should be $\sim10^5
\Msun$ year$^{-1}$.

\subsection{The Distance to \LkHa 233, Revisited}

\label{distance}

As discussed above, the fact that \LkHa 233 is seen entirely in
scattered light implies that it must be substantially more luminous
than has previously been thought, assuming it is indeed at a distance
of 880 pc. It is worth considering whether it might instead be closer.

Establishing accurate distances to Herbig Ae stars is often
problematic \citep[e.g.][]{1998AJ....116..890S,2004A&A...419..301M,Perrin2006PDS144}.
The 880 pc distance to \LkHa 233 was originally derived by \citet{1978MNRAS.182..687C}, and 
the majority of subsequent authors have adopted their estimate.
Calvet \& Cohen derived that value from the distance modulus to the 
nearby star HD 213976, a bright B1 star located 6\farcm5 away 
on the edge of the dark cloud in which \LkHa 233 is embedded.
They
also estimated photometric distances to two other members of this small cluster,
the K stars \LkHa 231
and 232, finding $d=600-700$ pc subject to the assumption that those stars are both
overluminous by 2 mags.\ in $V$ due to their youth. 

Is it possible that HD 213976 is in fact not at the same distance as
\LkHa 233? Aspin (1985) noted that HD 213976's optical colors imply its $A_V$ is only 0.42, significantly
less than the typical $A_V=2.5$ per kpc. This implies both  
that we look toward it along a galactic sightline with lower than average extinction, and 
that HD 213976 must be located on the near side of the dark cloud.
Hence \LkHa 233 should be at least as far away as HD 213976, if not
farther.

Some authors \citep[e.g.][]{2003MNRAS.340.1173B} have instead described \LkHa 233 as being 
a member of condensation A of the Lac OB1 molecular cloud, 
which is only $418\pm 15$ pc distant \citep{2005AJ....129..856H}.
While \LkHa 233's position on the sky does place it on the apparent edge of 
Lac OB1, we consider it unlikely to actually be a member of that association, for several reasons. First, \LkHa
233's proper motion is very different from that of Lac OB1: $(\mu_l \cos b, \mu_b) = (-9, 21)$ mas/yr
\citep{2005A&A...438..769D} versus
$(\mu_l \cos b, \mu_b) = (-2.3\pm0.1, -3.4\pm0.1)$ for Lac OB1;
\citep{1999AJ....117..354D}. 
Even more problematic is the fact that Lac OB1 is believed to be 16 Myr old
\citep{1999AJ....117..354D}. Because protoplanetary disks dissipate on
time scales of 3-10 Myr, the presence of a massive disk around \LkHa
233 is inconsistent with its being that old. 
For these reasons we prefer to retain the traditional 880 pc distance
for \LkHa 233. 

Because the system's inclination can be constrained based on the disk's
appearance, the proper motion of the clumps in the jet can potentially provide a 
direct measure of the distance. For a disk inclination $i=65\degr\pm 5$ and $d=880 \pc$,
the jet proper motion should be $0\farcs06\pm0.01$ year$^{-1}$, while
$d=420 \pc$ would imply a proper motion of $0\farcs12\pm0.03$ year$^{-1}$.
However, this will be a challenging measurement given uncertainty in the inclination,
the possibility of jet precession, variations in the internal working
surfaces traced by the clumps, etc.

\section{Discussion}

The outflow from \LkHa 233 is in general quite similar to
those seen around T Tauri stars. The velocity structure and collimation of
the jet, the morphology of the knots, and the presence of asymmetries
between the jet and counterjet all resemble the properties of outflows
seen around lower-mass stars. 

For instance, we see that for both the red and blue
directions, the higher velocity channels are more tightly collimated
than the lower velocity channels. Such increased collimation at higher
velocities has also been seen in the outflow from DG Tau
\citep{2002ApJ...576..222B}. This is in agreement with predictions of MHD models that
inner streamlines should dominate the emission
\citep[e.g.][]{2004Ap&SS.292..643D}.

It is striking how similar \LkHa 233 appears to the T Tauri star HH
30, with its well-known edge-on disk and perpendicular outflow
\citep{1996ApJ...473..437B}.  \LkHa 233's disk is
asymmetric in brightness between the left and right sides, as is 
HH 30's disk \citep{2004ApJ...602..860W}. The outflows
from both stars are knotty, and even the characteristic periods for knot
production are similar, 2.5 years for HH 30 and $\approx 5$ years for
\LkHa 233.
These knots appear to be due to internal shocks within the
jet, rather than the jet shocking into the ambient medium.
The properties of the knots in \LkHa 233's jet are indeed consistent with internal
working surfaces caused by variations in jet outflow velocity. In
particular, the brightest knot is also that with the highest velocity,
just what would be expected for more rapid material plowing into
a slower portion of the jet.  

It is curious that, at our angular
resolution and S/N, only the redshifted jet appears knotty.
Similar asymmetries have been seen in outflows from about
half of T Tauri
stars \citep{1994ApJ...427L..99H}.
On the other hand, at
least some jets are very symmetric, such as HH 212 in Orion
\citep{1998Natur.394..862Z}, which displays a very regular series of knots
in nearly perfect pairs on opposite sides of the star. The symmetry of
the HH 212 knots and the regularity of knot spacing in many
systems (including \LkHa 233) suggest that the production of knots is
an inherent part of the jet launch process, possibly due to disk
instabilities or recurrent stellar activity cycles. However, the
commonness of asymmetric systems indicates that any inherent
regularity in the process can easily be disrupted or masked, by
collisions with surrounding ambient material, pressure gradients, or asymmetries in
magnetic fields very near the star 
\citep{2006ApJ...650..985W}. But then what causes those asymmetries in
magnetic field or pressure?
The origin of these various symmetries and asymmetries remains mysterious, and poses a key challenge to 
theories of jet acceleration and collimation.

A particularly useful comparison object is 
HD 163296, which has perhaps the best-studied outflow from any Herbig Ae star
\citep{2000ApJ...542L.115D,2006ApJ...650..985W}. HD 163296, an A2Ve star at $122_{-13}^{+17}$
pc \citep{1998A&A...330..145V}, launches a bipolar collimated
jet, HH 409, which also is broken into an asymmetric series of knots. The HH 409 jet displays knots on both the jet
and counterjet, but on the two sides the knots differ in both and distance
from the star. Based on their high spatial resolution observations of HD 163296, Wassell et 
al.\ concluded that the asymmetries must originate near
or within the jet launch region itself, rather than being the result of
interactions with the ambient medium. Other authors have reached
the same conclusion regarding the asymmetries observed in outflows
from T Tauri stars \citep{2002ApJ...580..336W,2006astro.ph..6605P}.
Similarly for \LkHa 233 , given that strong
asymmetries are present in the outflow within the innermost few hundred
AU, it seems 
unlikely that those asymmetries could be due to the
ambient medium. While there is indeed substantial dust and gas
near the star (in the observed disk and nebula), we know the outflow
from \LkHa 233 has been ongoing for thousands of years based on
the length of the observed outflow (at least 1.3 pc; McGroarty et al.\
2004).  Therefore any primordial material along the jet axis was
long ago swept away, out of the inner few arcseconds where we observe
the asymmetries and knots.  Explaining the difference in appearance
between the red- and blue-shifted jets thus requires the presence of
an asymmetric geometry within the inner jet launch region. This
asymmetry may take the form of pressure gradients, different magnetic field strengths or differing field geometries
between the two directions.

The mass flux in the jet of \LkHa 233 is about two
orders of magnitude greater than that of HD 163296. The measured
electron density $n_e$ for \LkHa 233 is also about a factor of ten higher 
than that reported for HD 163296, but the value for HD 163296 was
measured using ratios of optical \SII transitions which trace lower density parts of
the post-shock gas \citep{2006ApJ...650..985W}.  Hence it is hard to say how meaningful this difference
in $n_e$ is, and the true difference in accretion
rate may be somewhat less than this estimate. 

Outflow mass fluxes are believed
to be proportional to accretion rate, and accretion rates are
generally larger
for younger stars \citep{1998ApJ...495..385H}.
Therefore, together with its
higher extinction, greater infrared excess, and larger amount of nebulosity, 
the high mass flux of \LkHa 233 suggests that it is at a
younger evolutionary state than HD 163296. This is consistent
with our revised stellar
parameters which place \LkHa 233 near the stellar birthline for
intermediate mass stars.
The outflow velocity is expected to be
of order the escape velocity, $v_{esc} = \sqrt{GM_\ast/R_\ast}$. 
For
our revised stellar properties (4 \Msun, 9 \Rsun), the escape velocity
is $v_{esc} \approx 300$ km s$^{-1}$, which compares favorably with the estimated total jet velocity.

There is a growing body of evidence that outflows from high mass stars
are generally less collimated than those of low mass stars. Given the
tightly collimated nature of \LkHa 233's outflow, the transition to
poorly collimated jets appears to occur at a higher mass.
However, the opening angle of the bipolar cavity in the envelope around \LkHa 233 \citep{Perrin2004Sci} 
is 30\degr, much
wider than the jet collimation angle. This, plus the scatter in HH objects
seen by McGroarty et al. (2004), argues either for a change in outflow direction with
time (such as from precession), for a much lower degree of
collimation during earlier stages of the outflow, or for the presence
of a wide-angle wind around the inner jet, unseen in the 
\FeII emission which traces relatively dense gas.

\section{Conclusions}

We have used Keck Observatory's laser guide star adaptive optics system and integral field
spectrograph OSIRIS to observe the bipolar outflow of the Herbig Ae
star \LkHa 233. Based on these observations, we believe that:

\begin{enumerate}
\item {\LkHa 233 has a narrow, collimated bipolar outflow, with the
      blueshifted jet at a position angle of 249\degr\ and the
      redshifted counterjet at 69\degr. }
\item{At the present angular resolution and sensitivity, the blueshifted 
    and redshifted jets have similar opening angles
      but very different appearances, with the red-shifted jet broken
      up into several knots while the blue-shifted jet is smooth. The
      knots in the red-shifted jet are consistent with shocked inner
      working surfaces within the jet.}
\item {Based on \FeII emission intensity, we estimate the mass flux in this jet to be $(1.2 \pm 0.3) \Exp{-7}
      \Msun$ year$^{-1}$. This is toward the top end of mass fluxes
      observed around other T Tauri and Herbig Ae stars, and suggests
      that \LkHa 233 is a very young source with strong ongoing
      accretion.}
\item{The observed properties of \LkHa 233's outflow -- its kinematics, collimation, 
      asymmetric and knotty appearance, electron density, and estimated mass flux -- 
      are all largely consistent with the observed properties of jets
      from T Tauri stars.  The transition to
      the less-collimated outflows observed around massive YSOs must
      therefore occur at some still higher mass threshhold.}
\item {The jet extends perpendicular to a flattened circumstellar
      disk, the innermost region of the X-shaped nebula around \LkHa
      233.} 
\item{\LkHa 233 is seen only in scattered light from this disk;
      the star itself remains obscured at near-infrared wavelengths. Because
      of this, previous
      estimates of the extinction to the star are low, causing the
      total stellar luminosity to be underestimated in prior studies.}
\item{Revised stellar parameters, based on fitting disk models to the
      SED, indicate that \LkHa 233 may be a 4 \Msun\ star near the
      stellar birthline, but there remain large uncertainties in its true
      stellar properties.}
\end{enumerate}

The main result of this study is confirmation that the outflow from
\LkHa 233 is quite similar to typical outflows from T Tauri stars, as
has been suggested based on prior observations at lower spatial
resolution \citep{corc98,2004A&A...415..189M}.  This supports the
hypothesis that the same physical mechanism is responsible for
creating stellar outflows, across the entire mass range from T Tauri
to Herbig Ae stars. It is worth noting that circumstellar disks
around Herbig Ae stars are also similar in broad outline to those
around T Tauri stars \citep[e.g.][]{grady2005,Perrin2006PDS144,2007prpl.conf..539M} In contrast, there
are suggestions that circumstellar disks around the Herbig Be
and more massive YSOs differ from those around lower masses. In particular, there
is evidence for a transition from magnetic accretion (for T Tauri and
Herbig Ae stars) to direct disk accretion onto the star (for Be stars
and beyond) \citep{2002MNRAS.337..356V,2007MNRAS.377.1363M}. Neither
this transition in accretion mechanisms nor the transition from
collimated to wide-angle outflows are particularly well understood.
But the fact that both transitions occur at approximately the same
stellar mass may be an important clue that these processes are
connected.  This may indicate either that the properties of outflows
depend primarily on the properties of the circumstellar disks and how
they accrete onto the star, or that both outflow and disk properties
depend in turn on some third parameter (stellar magnetic fields?)
which undergoes a transition at the Ae/Be boundary.

Further insight into these matters will depend on obtaining
increasingly detailed spatially resolved spectra of outflows, such as
those provided by integral field spectroscopy.  We have shown here the
measured electron density using the 1.644/1.600 $\mu$m ratio, and have
also presented a rough estimate of the mass flux using the 1.644
$\mu$m line. But much more could be done with a more complete
spectrum: Ratios of other lines (most notably 1.644/1.533) provide
alternate
measures of electron density, and comparison with \textit{optical}
\FeII lines can measure the temperature as well
\citep{2005A&A...441..159N,2006A&A...456..189P}.  The 1.644, 1.320,
and 1.257 \um\ lines all originate in the same upper state, and thus
their ratios directly constrain the extinction $A_V$.  This is a
particularly exciting prospect for the redshifted jet in \LkHa 233: it
is located partially behind the inclined circumstellar disk (occulting
it in the visible but not in the near infrared), and thus it can be
used to measure the extinction \textit{through the disk}.  Future
observations that include the complete optical and near-IR spectrum of
\FeII and other shock tracers should enable a complete suite of
diagnostics for temperature, density, ionization, refractory element
depletion, and more.  Detailed modeling of \LkHa 233 thus holds great
promise for understanding outflow mechanisms around intermediate-mass
stars.

\acknowledgments

These observations were made possible by the tremendous efforts of the
OSIRIS instrument team and Keck Observatory staff, whose members are
too numerous to list here in full. Special thanks in particular go to James
Larkin, Shelley Wright, Mike McElwain, Randy Campbell, and Al Conrad.
Conor Laver assisted with the OSIRIS data reduction pipeline. Christophe Pinte and
Gaspard Duchene provided excellent guidance in the use of their
MCFOST Monte Carlo code. MDP thanks Deirdre Coffey and Catherine
Dougados for enlightening discussions and warm hospitality during his
visit to Grenoble.  We thank the referee for thoughtful comments that
greatly improved this paper. 
The authors wish to recognize that the summit of Mauna Kea has always
held a very significant cultural role for the indigenous Hawaiian community.  
We are most fortunate to have the opportunity to observe from this mountain.

    This work has been supported in part by the National Science
      Foundation Science and Technology Center for Adaptive Optics,
      managed by the University of California at Santa Cruz under
      cooperative agreement No. AST-9876783.
      MDP was partially supported by a
      NASA Michelson Graduate Fellowship, under contract to the Jet
      Propulsion Laboratory (JPL). JPL is managed for NASA by the
      California Institute of Technology.

{\it Facilities:} 
\facility{Keck:II}

\bibliographystyle{apj-short}
\bibliography{haebes,pol}

\end{document}